**A Mixture of Experts (MoE) model to improve AI-based computational pathology prediction performance under variable levels of histopathology image blur**

Authors

Yujie Xiang[1], Bojing Liu[1], Mattias Rantalainen[1*]

[1] *Department of Medical Epidemiology and Biostatistics, Karolinska Institutet, Stockholm, Sweden*

* *Correspondence to* mattias.rantalainen@ki.se

**Abstract**

**Background**

AI-based models for analysis of histopathology whole slide images (WSIs) are now common. However, image quality, particularly unsharp areas of WSIs, impacts model performance. In this study we investigate the impact of blur on deep learning models for WSI analysis. We propose a mixture of experts (MoE) strategy that mitigates the impact of unsharp areas in WSIs on classification performance by combining predictions from multiple expert models trained on data with varying levels of blur.

**Methods**

The study included hematoxylin and eosin (H&E) stained WSIs from 2093 breast cancer patients. Classification of histological grades 1 and 3 was used as a primary benchmarking case, and prediction of immunohistochemistry (IHC) markers (ER, PR, HER2) from H&E as a secondary case. The proposed MoE strategy to improve robustness against blur was evaluated in both a deep CNN models (CNN_CLAM, and MoE-CNN_CLAM) and a Vision Transformer-based histopathology foundation model (UNI_CLAM and MoE-UNI_CLAM). For each architecture, a baseline model was trained on sharp images, and multiple expert models were trained on tiles with added Gaussian blur at different levels. Model performance (area under the ROC curve) was evaluated under multiple levels of uniform blur, as well as in several simulated scenarios with a mixture of blur levels within the WSIs.

**Results**

Baseline model performance degraded with increasing blur for all evaluated architectures. Individual expert models trained on data with simulated Gaussian blur performed better on unsharp images compared to baseline models. The proposed MoE consistently outperformed its respective baseline models in simulation scenarios with various degrees of blur within WSIs. MoE-CNN_CLAM outperformed the baseline CNN_CLAM under moderate (AUC: 0.868 vs. 0.702) and mixed blur conditions (AUC: 0.890 vs. 0.875). MoE-UNI_CLAM outperformed the baseline UNI_CLAM model in both moderate (AUC: 0.950 vs. 0.928) and mixed blur conditions (AUC: 0.944 vs. 0.931).

**Conclusions**

Unsharp image areas are common in WSIs and impact prediction performance. The proposed MoE strategy provided equal or substantially improved prediction performance under all evaluated test scenarios. The proposed methodology has the potential to increase quality and reliability of AI-based pathology models in both research and clinical applications.

# Introduction

Computational pathology[1] leverages artificial intelligence (AI) for the analysis of hematoxylin and eosin (H&E) stained digital histopathology whole slide images (WSIs), and enables model-based solutions both for routine pathology tasks and for precision medicine[2]. WSIs are the core data modality in digital pathology. WSIs are generated by scanning and digitizing a high-resolution pathological image of a tissue sample[3], and can subsequently be used both for manual inspection by a pathologist, and analysed by computer models. Deep convolutional neural networks (CNNs) are commonly used in computational pathology, which offers both computational efficiency and good performance[4] in routine pathological tasks of cancer detection and grading[2], and in predicting complex precision medicine tasks such as prognostic patient stratification[5] and prediction of treatment response[2]. More recently, Vision Transformer (ViT) with attention-based tile-to-slide aggregation models[6,7] is applied in computational pathology, like clustering-constrained-attention multiple-instance learning (CLAM)[7] framework. Foundation models (FMs) based on the ViT architecture[8–11] pre-trained on large-scale pan-cancer datasets, such as UNI[8], can potentially serve as universal frozen feature extractors and be further fine-tuned with attention-based architectures for task-specific applications.

Despite the increasing implementation of digital pathology in clinical and research settings, WSI quality is rarely perfect and uniform in real-world scenarios. In particular, there is often quality variability with respect to focus in different regions within a WSI, leading to the presence of areas with varying degrees of blur[12,13]. Image blur, commonly resulting from the scanning processes for histopathology slides[12,14], can adversely affect the performance of deep learning models in the analysis of WSIs[15–17]. This reduced performance is particularly undesirable in clinical applications that require high accuracy under real-world conditions.

Mixture of Experts (MoE) provides a modelling approach that enables the use of several specialised AI/ML models to improve flexibility in modelling and performance, with the capacity for handling heterogeneous or complex data distributions[25,26]. MoE enables training of multiple expert models to specialise in different subsets or aspects of the input data, a gating mechanism (or router) to dynamically select the most relevant expert(s), and a combination function that combines the contribution of each expert for each input[27–29] into a final prediction. The MoE approach has previously been applied in domains such as natural language processing[29,30], computer vision[25,31], and multimodal learning[32,33], where data heterogeneity and complex input structures are common. However, to our knowledge, MoE has not yet been systematically explored to address model image quality variations in AI-based pathology, specifically for mitigating the impact of blur in WSIs.

Motivated by the evident common presence of blurry areas in WSIs and the established negative impact on deep learning models, we hypothesise that the MoE approach in managing varying levels of blur at the stage of modelling can improve prediction performance. In this study, we first investigated the impact of varying image sharpness based on three commonly used model architectures: (1) CNN backbone with a simplified 75th-percentile[34] tile-to-slide aggregation (CNN_simple); (2) CNN feature extractor with CLAM attention-based tile-to-slide aggregation (CNN_CLAM); and (3) UNI FM feature extractor and CLAM attention-based tile-to-slide aggregation (UNI_CLAM). For each model architecture, we trained a baseline model with high-quality (sharp) WSI tiles, along with a set of experts with each trained on tiles augmented with a specific level of Gaussian blur, enabling specialisation for different quality conditions. At inference, a MoE strategy with a gating function or router based on an objective sharpness measure, assigns each image tile to the most appropriate expert for prediction. The final prediction for each WSI is obtained by combining the outputs from these selected experts, aiming to offer a more robust prediction in the presence of under-focused image regions.

In this study, we use a binary classification task of distinguishing between breast cancer Nottingham Histological Grade (NHG)[21,35] 1 and 3, as a principal example to illustrate the methodology. Our main contributions are in two areas: (1) the assessment of how CNN_simple, CNN_CLAM and UNI_CLAM models are affected by various degrees of blur, and (2) a MoE modelling strategy with a sharpness-based gating mechanism to dynamically route image tiles to specialised experts trained at different blur levels. This approach can improve model performance across heterogeneous image quality conditions in WSIs in both clinical and research settings. See Figure 1 for an overview. To further validate the effectiveness of MoE, we also evaluate the prediction of routine immunohistochemistry (IHC) biomarkers of estrogen receptor (ER), progesterone receptor (PR), and human epidermal growth factor receptor 2 (Her2) status from H&E WSIs.

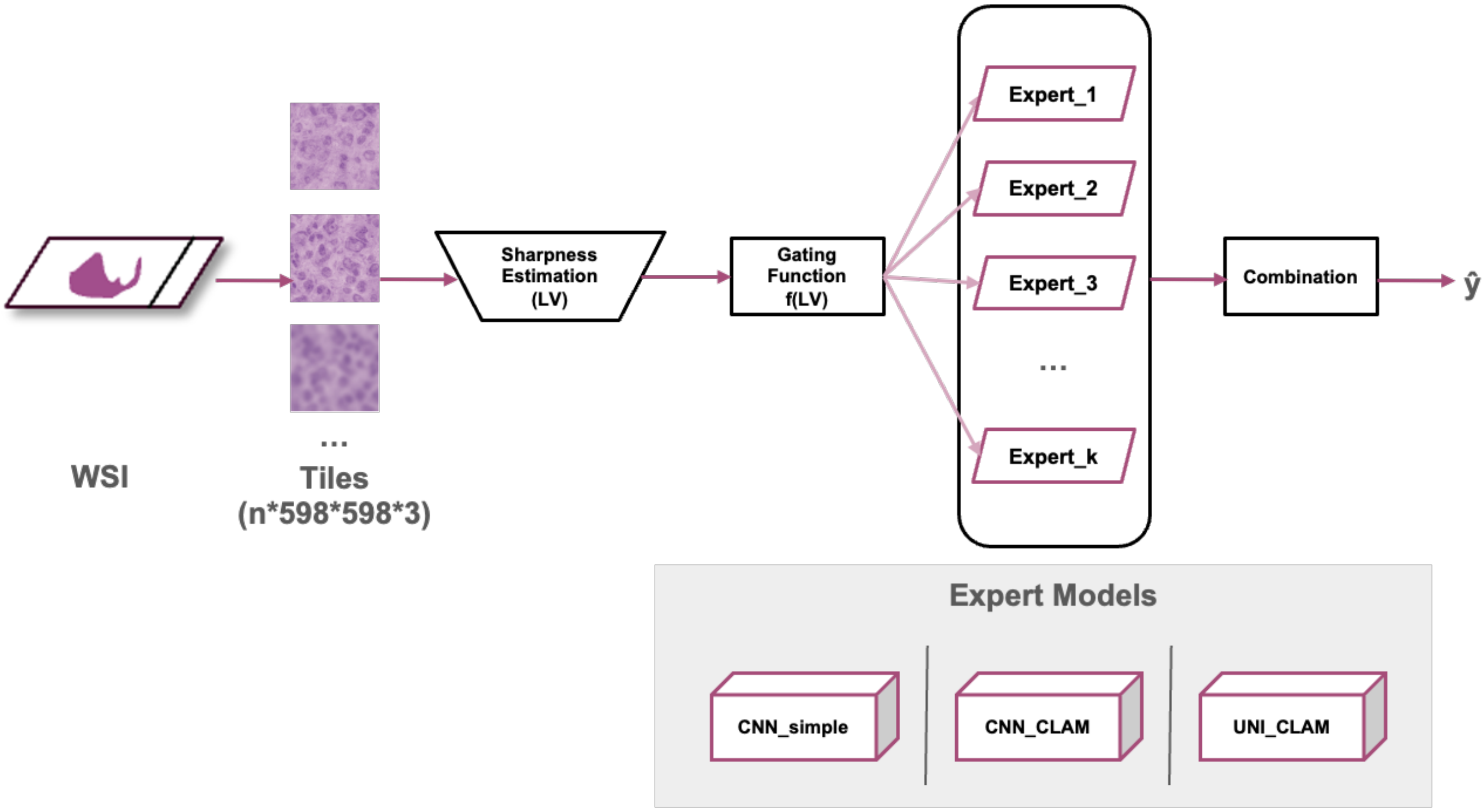

Figure 1. The workflow for the MoE strategy under heterogeneous blur conditions within WSIs. A WSI is divided into tiles. For each tile, sharpness is estimated using the variance of Laplacian (LV) to quantify local blur. This sharpness score is used by a gating function to assign the tile to the most appropriate expert model, each expert specialised for a different range of blur levels. Three types of expert models were evaluated, including CNN_simple, CNN_CLAM, and UNI_CLAM. Expert models independently predict tile-level outputs, which are subsequently combined (e.g., by the 75th percentile for MoE-CNN or by weighted averaging for MoE-CNN_CLAM and MoE-UNI_CLAM) to produce the final WSI-level prediction ($\hat{y}$). The entire process is repeated for each architecture. This MoE approach enables improved predictions across diverse image qualities found within a WSI.

# Materials and Methods

## Dataset

This study was based on data from a Swedish breast cancer cohort of female patients (N=2093). This cohort incorporated patients diagnosed at Stockholm's South General Hospital between 2012 and 2018, for whom archived histological slides and corresponding NHG grade were available[22]. The study has approval from the Swedish Ethical Review Authority (2017/2106-31, with amendments 2018/1462-32 and 2019–02336). No additional informed consent was required in accordance with ethical approval in this non-interventional collection and analysis of data from patient records. The primary analysis focused on distinguishing NHG 1 versus NHG 3 tumours, excluding patients with NHG 2 due to its ambiguous clinical significance, or those missing in NHG. The final dataset for NHG 1 vs. 3 classification included 916 patients (363 NHG 1, 553 NHG 3) (with one WSI per patient). Data were split into 5-fold cross-validation (CV) sets (80% training and 20% validation) at the patient level, with balanced distribution of NHG 1 and 3 cases across folds. For model selection and parameter tuning, we further divided each CV training set at the patient level into 80% for training and 20% for tuning (Supplementary Table S1), stratified by NHG.

In the secondary performance evaluation, focusing on IHC status classifications of ER, PR, Her2 from H&E slides, the final datasets consisted of 1656 (1483 ER-positive vs. 173 ER-negative), 1642 (1170 PR-positive vs. 472 PR-negative), and 1589 (176 Her2-positive vs. 1413 Her2-negative) patients, after excluding patients missing in each marker. Data splits for these tasks followed the same 5-fold CV and tuning set strategy (Supplementary Tables S2–S4).

# Image Pre-processing

We applied the same pre-processing and quality control steps to H&E-stained WSIs, as described previously [21,22]. In brief, the pre-processing steps comprised: (1) generating tissue masks using an Otsu threshold of 25 to exclude background regions, (2) tiling WSIs into 598 × 598 pixel image patches at 20X resolution (pixel size = 271 × 271 μm), (3) calculating the variance of Laplacian (LV) as a measurement of sharpness for each image tile, and discarding those with LV less than 500, indicating blurry or being out-of-focus (OOF), (4) applying a modified stain normalisation method proposed by Macenko et al.[36] across WSIs, and (5) applying a previously trained CNN model to detect cancer regions[21]. These pre-processed tiles were subsequently used in all model optimisation and analyses described below.

# Simulation of unsharp images using Gaussian blur

To simulate unsharp images, we used the Gaussian Blur function, also known as Gaussian Smoothing, to add and modulate the blurring intensity in the pre-processed tiles[37]. The function operates by averaging neighbouring pixels based on weights from a Gaussian distribution (i.e. Gaussian kernel)[38,39]. In this way, we can systematically investigate the impact of varying degrees of blur on model performance.

The Gaussian blur kernel is defined as follows (equation 1):

$$\mathbf{G}(x,y) = \frac{1}{2\pi\sigma^2} e^{-\frac{x^2+y^2}{2\sigma^2}}$$ [40]

where $G$(x,y) smooths the image by averaging neighbouring pixels, simulating image blur in WSIs, and x and y represent the spatial position relative to the centre of the kernel, and σ is the standard deviation of the Gaussian distribution, with larger σ resulting in a broader spread of the Gaussian kernel and a greater blur[41].

The modified (blurred) image is defined as follows (equation 2):

$$I'(x, y) = (I * G)(x, y) \quad ^{42}$$

where $I'(x, y)$ denotes the blurred image resulting from the convolution of original image $I$ with Gaussian kernel $G$[42,43].

# Sensitivity analysis of blur on single model performance

We hypothesised that a model trained on blurry images might outperform one trained on non-blurry images when predicting on unsharp images. To test this hypothesis, we simulated varying levels of blurriness on WSIs both for model training and performance evaluation.

## Training of models on varying degrees of unsharp images

First, we created ten blurred training sets, where all tiles in the training sets were transformed by the addition of a Gaussian blur with $\sigma \in \{0.5, 1.0, 2.0, 3.0, 4.0, 5.0, 6.0, 7.0, 8.0, \text{and } 9.0\}$. These blurred data were used to train the basic CNN_simple models, additional attention weights for CNN_CLAM models, as well as attention weights for UNI_CLAM models. The UNI FM was used as a frozen feature extractor without fine-tuning on varying blur levels.

## Evaluation of model performance on varying degrees of unsharp images

To validate the performance of these models on images with varying degrees of blur, we simulated 14 blurred validation sets, each with added Gaussian blur at different sigma levels, from subtle to significant: 0.5, 1.0, 1.5, 2.0, 2.5, 3.0, 3.5, 4.0, 5.0, 6.0, 7.0, 8.0, 9.0, and 10.0 (Figure 2). A sigma of zero indicated no added blur (i.e. the original focused tiles). These data were used to assess the sensitivity of different models to varying degrees of blur, and to establish cut-points for each expert in our MoE framework. Original image tiles, after excluding the highly blurry tiles, were considered sharp (see above). With increasing sigma values, blur effects gradually intensified on the image tiles. For instance, at lower levels (e.g., $\sigma \leq 1.5$), certain morphologies remained visible, but with sigma values of 5.0 to 10.0, most microscopic morphological structures were smoothed out and no longer distinguishable (Figure 2). Based on these observations, we stratified the blurred tiles into three levels, i.e., low blur ($\sigma \in [0.0, 1.5]$): minimal blur with preserved morphological features, moderate blur ($\sigma \in (1.5, 5.0)$): noticeable blurry but not a complete loss of detail, and high blur ($\sigma \in [5.0, 10.0]$): severe blur with most morphology lost.

## Benchmarking and performance metrics

The prediction performance of all three model architectures was systematically evaluated using 5-fold CV on data with varying levels of simulated Gaussian blur, including the original tile set and 14 tile sets with incrementally increased levels of simulated Gaussian blur (15 levels in total, see Figure 2). For the binary classification of NHG 1 vs. NHG 3, we calculated and compared the area under the ROC curve (AUC). For each validation set, the CNN_simple model was evaluated directly on tile images, while the attention-based models (CNN_CLAM and UNI_CLAM) were evaluated using feature vectors extracted by the corresponding backbone (CNN or UNI) from the same validation tiles.

# Mapping tiles with the simulated Gaussian blur to the Variance of Laplacian (LV)

We were interested in understanding how tiles with simulated Gaussian blur could reflect real-world images with different qualities. As LV serves as a quantitative index of sharpness and has been commonly employed in evaluating the quality of histopathology images[44,45], we established a mapping between different levels of Gaussian blur (i.e., with sigma ranging from 0.5 to 10) and LV (Figure 3). This mapping facilitated our understanding of how the simulated blurry tiles correspond to the LV levels in real-world histopathology images. First, for a representative assessment, we randomly selected 10,000 tiles from the original dataset. Second, we applied a series of sigma values to the selected original tiles to introduce blur. Finally, the LV was computed for each original tile and for each tile after applying various Gaussian blur. A higher LV indicates a sharper tile. The median LV value of these 10,000 tiles at each sigma is presented in Table S5 in the supplementary materials.

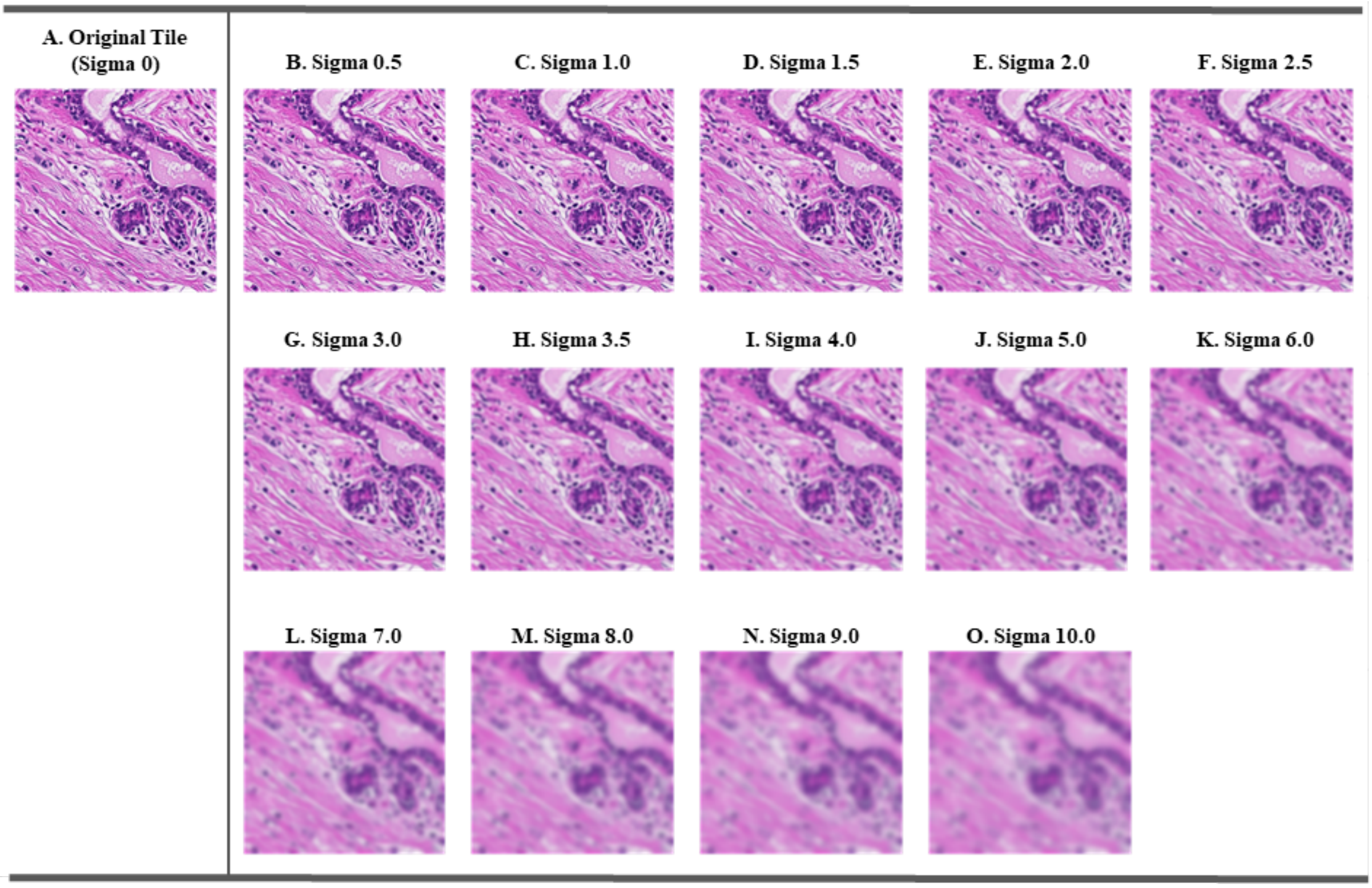

Figure 2. Visualisation of the impact of various levels of Gaussian blur in histopathology images tiles. Image (A) represents the original, unblurred image, serving as the baseline for the analysis. Images (B-O) show the application of Gaussian blur with increasing sigma values from 0.5 to 10.0. Specifically, these images illustrate the effect of varying degrees of blurring added to the original histopathology image.

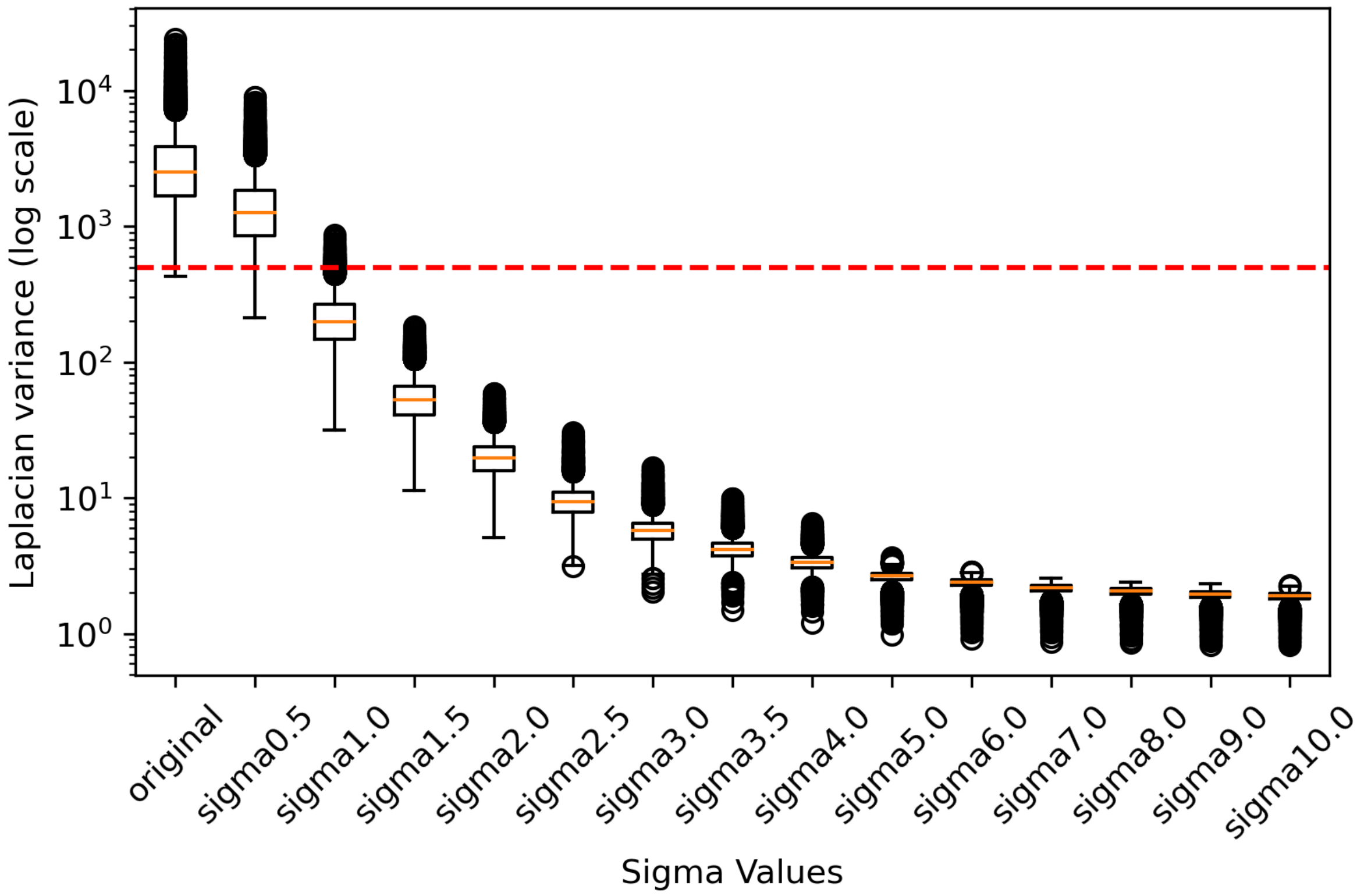

Figure 3. The variance of Laplacian for tiles with different Gaussian blur levels added. The y-axis shows the variance of the Laplacian values plotted on a logarithmic scale. The red line represents the LV threshold of 500, which is the cut-off used to filter blurry tiles in the original dataset during pre-processing.

# Optimisation of base models and models optimised for blurry image data

## Training of deep CNN - baseline and expert models

We trained a set of baseline and multiple expert models based on the CNN_simple architecture using tiles with varying levels of Gaussian blur. The baseline model (i.e. Model_Base) was trained with original image tiles (no added Gaussian blur) to classify NHG 1 vs. 3, using ResNet18 backbone with ImageNet initial weights[46]. The model was optimised using the Stochastic Gradient Descent (SGD)[47] optimiser with a cross-entropy loss function. We used a mini-batch size of 32 per step, and each optimisation partial epoch consisted of 1250 steps (including 40,000 tiles in total) for training. We randomly selected 625 mini-batches (20,000 tiles) in the tuning set to evaluate model performance at each epoch end. Early stopping was applied when the loss did not decrease after a patience of 25 partial epochs or continued training for a maximum of 500 epochs. Data augmentation with rotation and flips was applied during model training. The initial learning rate (LR) was set to 1e-5. A LR scheduler (ReduceLROnPlateau) was employed to reduce the LR by a factor of 0.2 when the validation loss plateaued, with a patience of 13 partial epochs. The models were trained using PyTorch 1.13.0 and RayTune 2.1.0 frameworks on four RTX 2080 Ti GPUs.

Ten additional expert models were trained on data augmented with varying levels of added Gaussian blur ($\sigma \in \{0.5, 1.0, 2.0, 3.0, 4.0, 5.0, 6.0, 7.0, 8.0, 9.0\}$), respectively. All expert models also used the same Resnet18 backbone as the above-mentioned baseline model. We optimised the LR for each model trained on data with varying levels of Gaussian blur ($\sigma$) as follows: Model_0.5 at 1e-5, Model_1.0, Model_2.0, Model_3.0, and Model_4.0 at 1e-6, and the rest models at 3e-7, while keeping other hyperparameters consistent with the Model_Base. For all baseline and expert models, we applied a tile-to-slide aggregation method using the 75th percentile of tile-level prediction.

## Training of deep CNN with attention mechanism - baseline and expert models

In addition, we applied the same Gaussian blur simulation and training strategy to the CNN_CLAM architecture, which combined a ResNet18 feature extractor (trained on the specific task) with an attention mechanism to aggregate information from tile to slide. For the attention module, we adapted a simplified CLAM[7] framework, which operates bag-level supervision only, without instance-level supervision. With this simplification, and in line with the principles of multiple instance learning, we hypothesised that a properly trained attention mechanism could enable the model to down-weight blurry tiles and assign greater importance to sharp tiles during aggregation, thereby improving performance on WSIs with blurred regions. Specifically, we trained CNN_CLAM baseline (i.e. using sharp tiles) and ten experts on tile sets with varying levels of simulated Gaussian blur ($\sigma \in \{0.5, 1.0, 2.0, 3.0, 4.0, 5.0, 6.0, 7.0, 8.0, 9.0\}$). Each model used a corresponding CNN_simple expert, previously trained on tiles with the same blur level, as a fixed feature extractor. For each tile, we obtained a 512-dimensional feature vector from the penultimate layer of respective CNN_simple expert. For each WSI, these tile-level features (from the same feature extractor) were fed into the attention module to optimise the attention weights. The CNN_CLAM models were trained to perform the same patient-level binary classification task (i.e., NHG 1 vs. 3 classification).

## Training of ViT Foundation Model with attention mechanism - baseline and expert models

In the third modelling approach, UNI_CLAM, we extracted image features using the pre-trained ViT-based FM, UNI[8] , and trained the attention module based on tiles with varying levels of simulated Gaussian blur (ten sets with $\sigma$ ranging from 0.5 to 9.0) for our specific benchmarking tasks. Specifically, for each set of original tiles and tiles with simulated Gaussian blur, we extracted the 1024-dimensional features from the UNI extractor. These features were then used to train the attention module (analogous to the CNN_CLAM approach) to perform the same patient-level binary classification tasks (NHG 1 vs 3, and IHC marker status classification).

CNN_CLAM and UNI_CLAM models were trained following the same training protocol. Each model was trained for a maximum of 100 epochs using the Adam optimiser, with an initial LR of 1e-5. A ReduceLROnPlateau scheduler was employed with a patience of 13 epochs, and early stopping was applied with a patience of 25 epochs to prevent overfitting. The mini-batch size was set to 24. We enabled dropout in the attention layers to promote generalisability. The training was conducted using 5-fold CV with stratified patient-level splits, using the same CV splits used for the CNN models to ensure directly comparable results. All training and evaluation were performed using tile-level feature vectors.

# Mixture of experts (MoE) model for blur-robust predictions

The proposed MoE method integrated multiple expert models, each specialised for a distinct range of image blur conditions within WSIs. The MoE framework consists of three primary steps during inference: (1) a tile-level sharpness (blur) estimation, (2) a gating mechanism that dynamically assigns each tile to the most appropriate expert model based on its estimated blur level, and (3) a combination function that aggregates results from the set of experts (Figure 1).

## Gating mechanism: blur estimation and expert assignment

Blur estimation was performed at the tile level and quantified using LV. Based on the LV value, a gating function assigned each tile to the corresponding expert specialised for its blur level. The domain boundaries for experts were empirically defined by mapping sigma-based blur levels to the corresponding LV thresholds (Supplementary Table S5). Specifically, we evaluated the performance of each expert model on tiles with simulated Gaussian blur across a range of sigma levels, allowing us to define blur-specific thresholds for expert selection. For each sigma-defined blur category, we computed the median LV for that sigma level and the next consecutive level, and averaged them to determine LV thresholds that separate blur categories. The gating function used these calculated LV thresholds to assign image tiles to their optimal expert models.

## Implementation across model architectures

This MoE strategy was implemented and tested in CNN_simple, CNN_CLAM, and UNI_CLAM. It was initially implemented using CNN_simple-based experts (MoE-CNN_simple) and subsequently extended to attention-based experts (MoE-CNN_CLAM and MoE-UNI_CLAM).

## Slide-level prediction aggregation and MoE combination strategy

The prediction at slide level of MoE-CNN_simple variant was aggregated from the 75th percentile of tile-level prediction. The final prediction for each WSI using the MoE-CNN_CLAM and MoE-UNI_CLAM variants was obtained via a weighted average approach (equation 3, 4 & 5).

The MoE model has *m* expert models. For each WSI, the weight assigned to the expert model $m_i$ ($w_{m_i}$) was calculated as the fraction of tiles handled by each expert model $m_i$ (equation 3):

$$w_{m_i} = \frac{n_{m_i}}{n_{tot}} ,$$

where $n_{m_i}$ is the number of tiles from the WSI assigned to the expert model $m_i$, and $n_{tot}$ is the total number of tiles for the WSI, and where (equation 4)

$$\sum_{i=1}^{m} w_{m_i} = 1 .$$

Let $\hat{p}_1 \ldots, \hat{p}_m$ be the predictions from the individual expert models. Then we define the MoE slide-level prediction ($\hat{p}_{MoE_final}$) as (equation 5):

$$\hat{p}_{MoE_final} = \sum_{i=1}^{m} w_{m_i} * \hat{p}_{m_i} ,$$

where $\hat{p}_{m_i}$ denotes the probability output by the $i_{th}$ expert model.

The use of a weighted average as the combination function allows us to account for the proportion of tiles allocated to each expert, such that the models with a larger number of tiles analysed will have a proportionally larger weight in the final prediction, while still incorporating information from all experts. See Tables 2 and 3 for the algorithm descriptions of the MoE strategy.

## Simulating varying Gaussian blur within WSIs in validation sets to emulate real-world data

To validate the MoE framework under realistic conditions, we simulated blurred validation sets mimicking the real-world data where WSIs have regions with heterogeneous focus qualities. Our simulation involved incorporating varying Gaussian blur for different proportions of tiles within a single WSI. This simulation was performed under 12 scenarios (Table 1). For the simulation purpose, we categorised the degree of Gaussian blur into three empirically defined levels based on observed changes in morphological detail at specific sigma values, i.e., low blur: $\sigma \sim U(0,1.5)$, moderate blur: $\sigma \sim U(1.5,5.0)$, and high blur: $\sigma \sim U(5.0,10.0)$. For each sigma group, the application of Gaussian blur on a single tile was determined by random sampling from a uniform distribution[48] defined by the corresponding sigma range. This approach ensured that the degree of blur applied to the tiles was both randomised and evenly distributed within predefined ranges. These simulated blurred validation sets were used to benchmark MoE performance against baseline single-expert models in WSIs with heterogeneous focus quality.

Table 1. Overview of the 12 simulation scenarios*

| Scenarios | Low blur % | Moderate Blur % | High blur % |
|---|---|---|---|
| 1 | 100 | 0 | 0 |
| 2 | 0 | 100 | 0 |
| 3 | 0 | 0 | 100 |
| 4 | 50 | 25 | 25 |
| 5 | 25 | 50 | 25 |
| 6 | 25 | 25 | 50 |
| 7 | 50 | 50 | 0 |
| 8 | 0 | 50 | 50 |
| 9 | 80 | 10 | 10 |
| 10 | 10 | 80 | 10 |
| 11 | 10 | 10 | 80 |
| 12 | 80 | 15 | 5 |

*12 simulation scenarios assigned varying proportions of tiles within a WSI to low blur, moderate blur, and high blur groups. These simulated datasets were used to validate MoE approach performance.

Table 2. Algorithm overview - MoE-CNN_simple

| For each tile within a WSI in each scenario: | |
|---|---|
| Step 1. | The initial blur level of a tile was denoted by $g$, representing the baseline sigma of the Gaussian blur. |
| Step 2. | Each tile within a WSI was randomly assigned to one of three predefined Gaussian blur groups (Low, Moderate, and High blur) according to the specific scenario listed in Table |
| | 1. Each group corresponds to a different range of blur severity. |
| Step 3. | A new blur value $g_i$ was applied to each tile based on its assigned Gaussian group using a uniform distribution, i.e. $g_i \sim U(a, b)$, where $a$ and $b$ define the sigma range of the assigned group. The final blur level for the tile was then computed as $\hat{g} = g + g_i$. |
| Step 4. | The LV $\theta$ of a newly blurred tile was calculated as: $\theta = LV(\hat{g}) = Var(L * I_{\hat{g}})$, where $L$ is the Laplacian operator, $I_{\hat{g}}$ is the simulated image with Gaussian blur $\hat{g}$, and $\theta$ is the estimated LV value for the blurred tile. |
| Step 5. | The gating mechanism assigned each tile to a specific expert (e.g., m CNN_simple models are included in MoE-CNN_simple) based on the estimated $\theta$ and pre-determined LV thresholds ($\tau$) derived from sigma-to-LV mapping(if $\theta > \tau_1$, expert_1 is assigned; if $\tau_2 < \theta \leq \tau_1$, expert_2 is assigned; if $\tau_3 < \theta \leq \tau_2$, expert_3 is assigned, …, if $\theta \leq \tau_{m-1}$, expert_m is assigned). Each expert produced predictions only for tiles assigned to it. |
| Step 6. | Steps 2 through 5 were repeated for all tiles within the WSI, until predictions were generated for every tile under the specified scenario. The prediction at slide-level for this WSI is aggregated from the 75th percentile of tile-level prediction. |

Table 3. Algorithm overview - MoE-CNN_CLAM or MoE-UNI_CLAM

| For each tile within a WSI in each scenario: | |
|---|---|
| Step 1. | The initial blur level of a tile was denoted by $g$, representing the baseline sigma of the Gaussian blur. |
| Step 2. | Each tile within a WSI was randomly assigned into one of three predefined Gaussian blur groups (Low, Moderate, and High blur) according to the specific scenario listed in Table 1. Each group corresponds to a different range of blur severity. |
| Step 3. | A new blur value $g_i$ was applied to each tile based on its assigned Gaussian group using a uniform distribution, i.e. $g_i \sim U(a, b)$, where $a$ and $b$ define the sigma range of the assigned group. The final blur level for the tile was then computed as $\hat{g} = g + g_i$. |
| Step 4. | The LV $\theta$ of a newly blurred tile was calculated as: $\theta = LV(\hat{g}) = Var(L * I_{\hat{g}})$, where $L$ is the Laplacian operator, $I_{\hat{g}}$ is the simulated image with Gaussian blur $\hat{g}$, and $\theta$ is the estimated LV value for the blurred tile. |
| Step 5. | The feature vector of tile is extracted by CNN (512 dimensions) or UNI (1024 dimensions). |
| Step 6. | The gating mechanism assigned the tile to a specific expert (e.g., m expert models are included in attention-based MoE) based on the estimated $\theta$ and pre-determined LV thresholds ($\tau$) derived from sigma-to-LV mapping(if $\theta > \tau_1$, expert_1 is assigned; if $\tau_2 < \theta \leq \tau_1$, expert_2 is assigned; if $\tau_3 < \theta \leq \tau_2$, expert_3 is assigned, …, if $\theta \leq \tau_{m-1}$, expert_m is assigned). Each expert produced predictions only for tiles assigned to it. |
| Step 7. | Steps 2 through 6 were repeated for all tiles within the WSI until predictions were |
| | generated for every tile under the specified scenario. The prediction for this WSI is generated using weighted averaging. |

## Additional benchmarking tasks

To further evaluate the generalisability of the MoE framework, we extended the analysis to three additional binary classification tasks, i.e., classification of ER, PR, and Her2 biomarker status from H&E WSI images. For each task, we trained classification models (UNI_CLAM_ER, UNI_CLAM_PR, and UNI_CLAM_Her2, respectively) using the same training pipeline, model architecture, and optimisation settings as described for the grade classification task (UNI_CLAM). The validation procedure and MoE modelling strategy were applied consistently to each biomarker classification task to assess model performance under varying levels of blur.

# Results

## Sensitivity towards blur on deep CNN model performance

We first evaluated the prediction performance of one baseline CNN_simple and ten blur-resistant CNN_simple models, each trained on tiles with a specific simulated Gaussian blur (σ ranging from 0.5 to 10), on the validation set consisting of sharp (without simulated blur) tiles exclusively. The AUCs for each model were as follows: 0.887 for Model_Base, 0.864 for Model_0.5, and 0.776 for Model_1.0, 0.750 for Model_2.0, 0.715 for Model_3.0, 0.682 for Model_4.0, 0.610 for Model_5.0, 0.508 for model_6.0, 0.581 for model_7.0, 0.544 for Model_8.0, and 0.496 for model_9.0, confirming that Mobel_Base that was trained on sharp images perform the best on unblurred data.

Next, we evaluated these 11 models across validation sets with increasing levels of Gaussian blur (Figure 4). We noted that models' performance depended on the levels of blur in the training data, showing varying degrees of robustness on validation sets with different levels of blur. For instance, models trained with mildly blurred tiles (e.g., Model_Base, Model_0.5) performed well on the original and slightly blurred data, but their performance decreased rapidly as the intensity of Gaussian blur increased. In contrast, models trained with more heavily blurred tiles (e.g., Model_3.0, Model_4.0, and Model_5.0) performed poorly on sharp tiles, but showed higher AUCs across moderate to high blur levels.

In Figure 4, Model_Base demonstrated superior performance compared to the other models under conditions of mild blur ($\sigma < 1.5$). In the range of $1.5 \leq \sigma < 2.5$, Model_0.5 outperformed all other models. The performance of Model_3.0 exceeded other models under moderate blur ($2.5 \leq \sigma < 3.5$), while Model_4.0 performed the best for higher blur levels ($3.5 \leq \sigma < 5.0$). Under conditions of extreme blur ($\sigma \geq 5.0$), Model_5.0 consistently yielded the highest AUCs. According to these findings, the sharpness levels were divided into five groups (Figure 4, indicated by coloured background regions), with each group corresponding to the best-performing model within that blur range. Next, we quantified the relationship between Gaussian blur (i.e. $\sigma$) and image sharpness as measured by LV (Figure 3). This allowed us to define LV-based thresholds for model selection in the MoE approach (Supplementary Table S6).

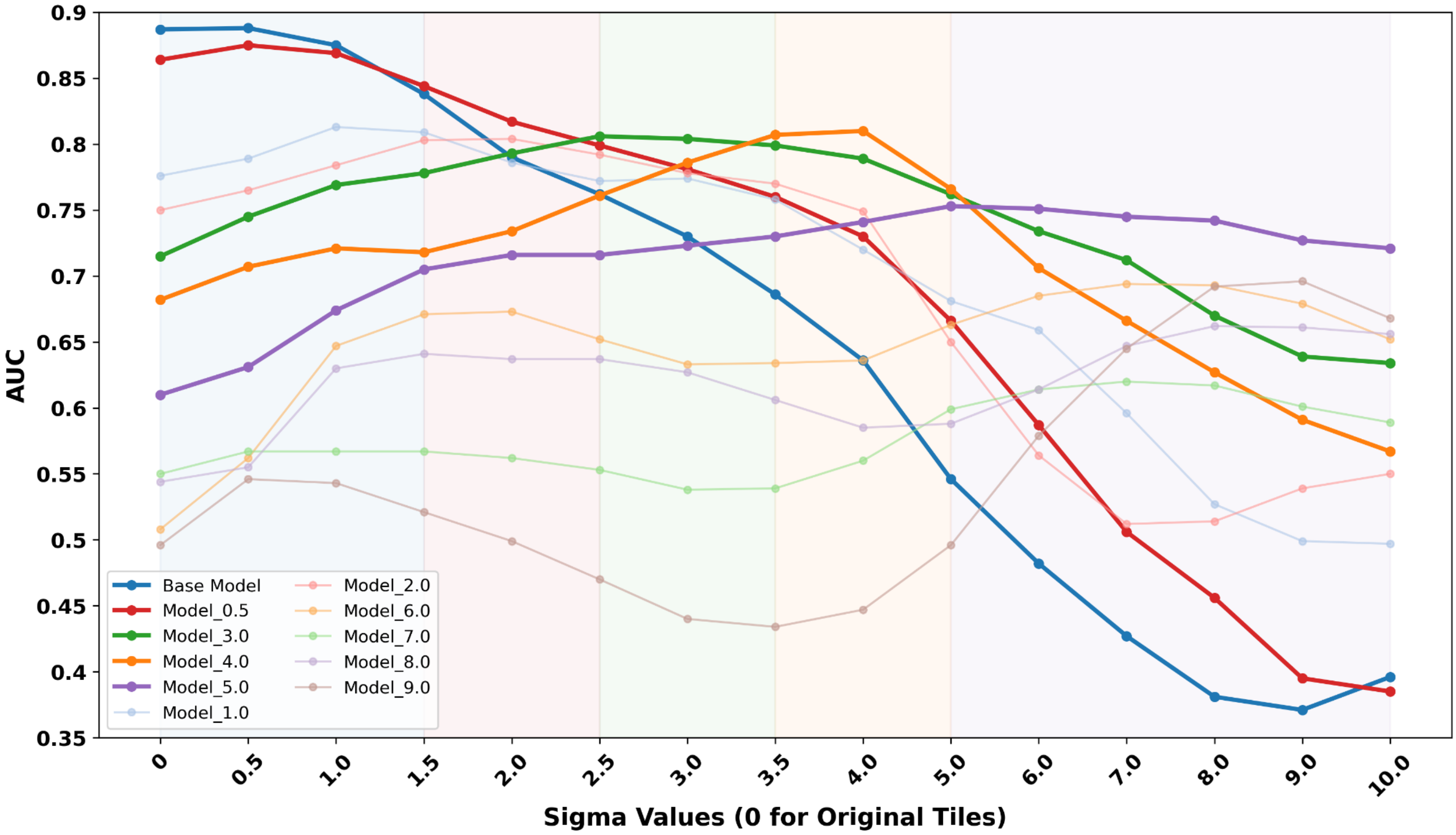

Figure 4. AUC performance of one baseline and ten blur-resistant CNN_simple models and cut-offs for model selection in MoE. Prediction performance (i.e. AUC) of one baseline and ten blur-resistant CNN_simple models evaluated on validation sets with increasing levels of Gaussian blur (σ ranging from 0 to 10). Model_Base was trained on the original (non-blurred) tiles, while the other models were each trained on tiles blurred with a fixed sigma value (e.g., Model_0.5 was trained on tiles with σ = 0.5). The x-axis represents the sigma values applied to the validation tiles, simulating an increasing level of blur. The y-axis shows the AUC of each model's prediction at the slide level. Coloured background regions denote the empirically defined blur category cut-offs used in MoE strategy for model selection.

# Sensitivity towards blur of the attention-based models based on CNN or ViT (Foundation Model) architectures

We evaluated the performance of a CNN model with attention mechanism (CNN_CLAM) and ViT foundation model with an attention mechanism (UNI_CLAM) across a series of image blur conditions.

## CNN model with attention mechanism results

As shown in Figure 5, the AUCs of the CNN_CLAM models varied across validation sets with increasing blur intensity (σ ranging from 0 to 10). Consistent with prior observations in CNN_simple models, CNN_CLAM models trained on tiles with lower simulated blur (e.g., CNN_CLAM_Base and CNN_CLAM_0.5) demonstrated high AUCs on sharp and mildly blurred validation sets, but their performance declined as blur level increased (i.e. $\sigma > 1.5$). Conversely, models trained on moderately blurred tiles (i.e., CNN_CLAM_2.0 to CNN_CLAM_5.0) maintained relatively stable performance across a broader range of blur levels compared to the CNN_CLAM_Base model.

Notably, almost every CNN_CLAM model achieved peak AUC when evaluated on tiles blurred at the level matching its respective training data. Furthermore, the best-performing model varied depending on the level of blur. Specifically, the baseline CNN_CLAM model achieved the highest AUC on sharper images ($\sigma < 1.5$), while models trained with increasing blur showed superior performance on increasingly degraded inputs. For example, CNN_CLAM $\sigma = 4.0$ and 5.0 performed the best under moderate-to-severe blur conditions. These findings support the motivation for the proposed MoE selection strategy, which selects the appropriate expert model based on the blur characteristics of the input tiles (Supplementary Table S6).

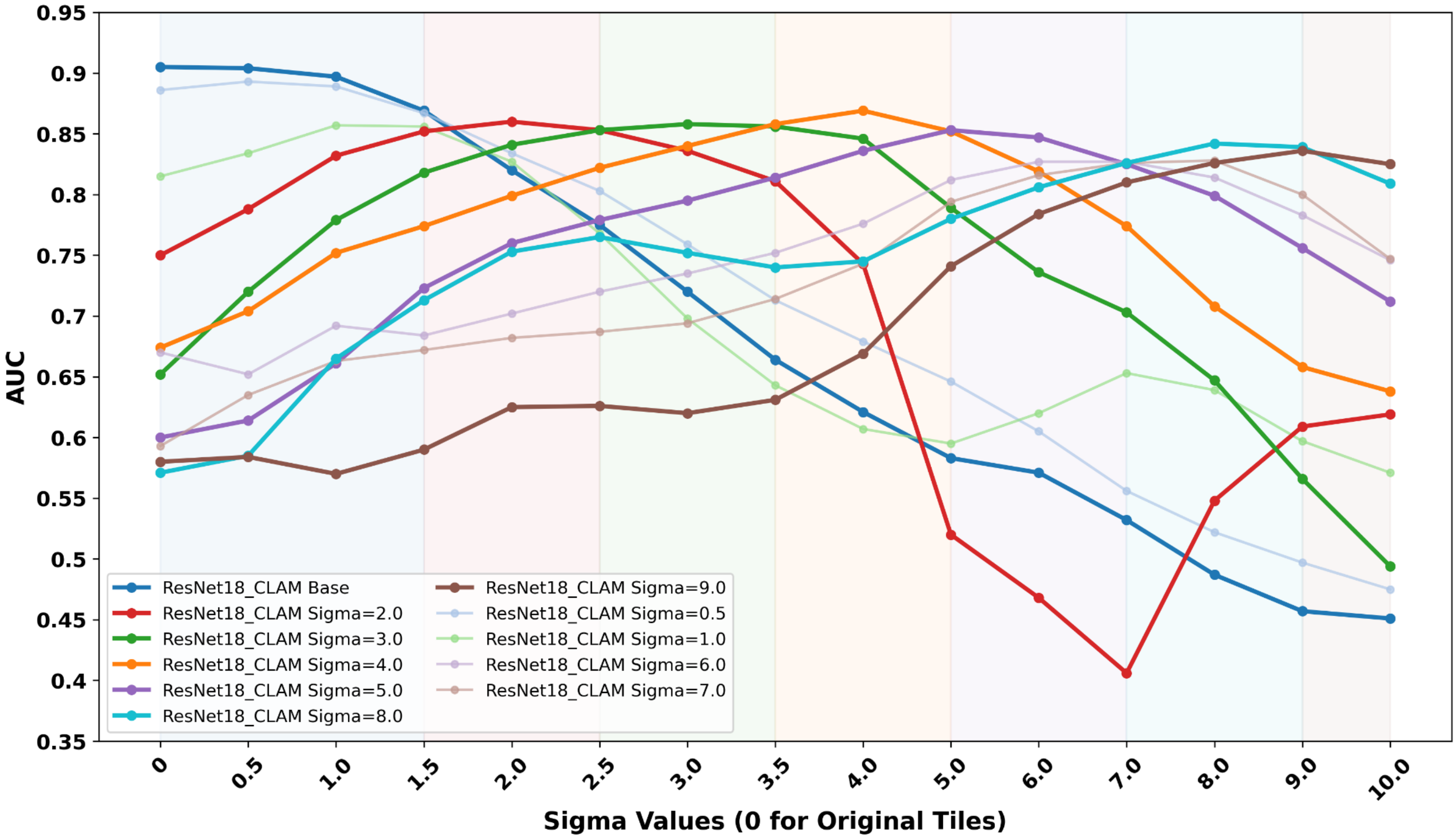

Figure 5. AUC performance of one baseline and ten blur-resistant CNN_CLAM models and cut-offs for model selection. AUC of one baseline and ten blur-resistant CNN_CLAM models evaluated on validation sets with increasing levels of Gaussian blur (σ ranging from 0 to 10). Each CNN_CLAM model was trained on features of tiles extracted from ResNet18 models trained on tiles with a fixed blur level. The x-axis represents the sigma values applied to the validation tiles, simulating an increasing level of blur. The y-axis shows the AUC of each model's prediction at the slide level. Coloured background regions denote the empirically defined blur category cut-offs used in MoE strategy for model selection.

## ViT foundation model with attention mechanism results

Figure 6 presents the performance of UNI_CLAM. Similar to the previous experiments, model performance was evaluated across validation sets with increasing blur intensity. These results revealed a distinct blur-specific performance pattern. Models trained on lower blur levels ($\sigma = 0.0$ to $\sigma = 3.0$) showed excellent performance on sharp tiles and tiles with low-to-moderate blur ($\sigma \leq 3$). However, their AUC scores gradually declined as the validation blur intensity increased. In contrast, models trained at higher blur levels (σ = 4.0 to σ = 9.0) underperformed on sharp and low-to-moderate blurred images but peaked near their respective training level, with their performance declined beyond that point. This demonstrated a strong specificity of UNI_CLAM models to their training blur levels and emphasised the difficulty of generalising a single attention-based model across diverse blur intensities. Based on the observed trends, the MoE selection strategy was defined for UNI_CLAM, guided by LV thresholds derived from validation blur levels (Supplementary Table S6).

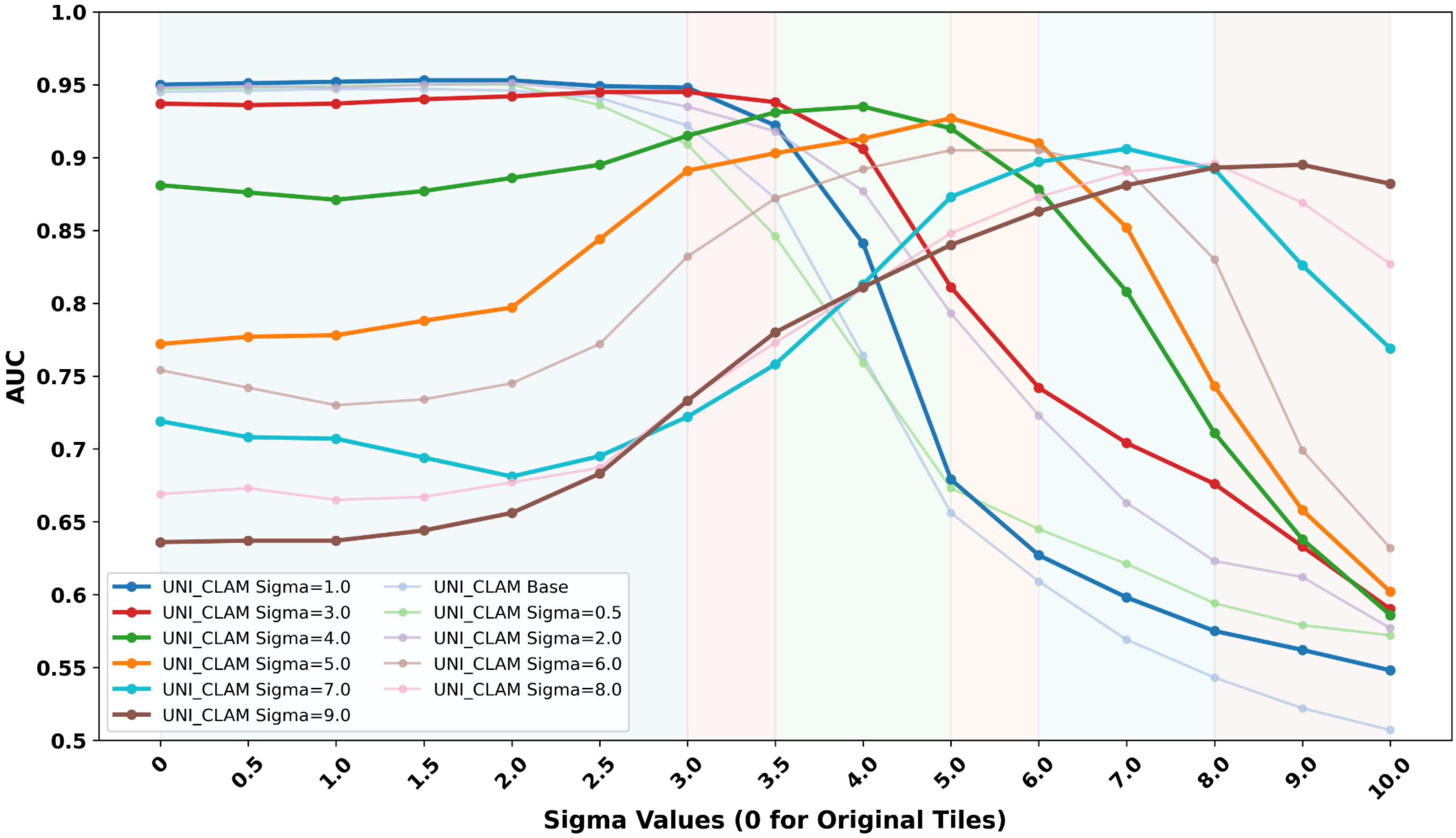

Figure 6. AUC performance of one baseline and ten blur-resistant UNI_CLAM models and cut-offs for model selection. AUC of one baseline and ten blur-resistant UNI_CLAM models evaluated on validation sets with increasing levels of Gaussian blur (σ ranging from 0 to 10). Each UNI_CLAM model was trained on features of tiles with a fixed blur level extracted using UNI. The x-axis represents the sigma values applied to the validation tiles, simulating an increasing level of blur. The y-axis shows the AUC of each model's prediction at the slide

level. Coloured background regions denote the empirically defined blur category cut-offs used in MoE strategy for model selection.

## Comparison of blur sensitivity of the proposed MoE model and baseline models

We simulated validation data across 12 different scenarios to mimic the real-world data with varying quality. We compared the model performance from the MoE (MoE-CNN_simple, MoE-CNN_CLAM, and MoE-UNI_CLAM) and the baseline model (CNN_simple_Base, CNN_CLAM_Base, and UNI_CLAM_Base) approaches (Table 4).

The MoE approaches consistently achieved higher AUC scores compared to their baseline models across all scenarios, except for scenario 1, where the performance of MoE and baseline model were on par. The most notable improvements were observed in scenarios where a substantial proportion of tiles were moderately or severely blurred (i.e. scenarios 2, 3, 8, 10, 11). For instance, in scenario 2 (100% moderate blurred tiles) and 3 (100% heavily blurred tiles), the AUCs for MoE-UNI_CLAM vs. UNI_CLAM_Base were 0.950 vs. 0.928 and 0.879 vs. 0.578, respectively. A similar trend was observed for MoE-CNN_simple and MoE-CNN_CLAM. In scenarios involving a combination of moderately and heavily blurred tiles, i.e. scenario 8 (50% moderate and 50% heavy), 10 (80% moderate and 10% heavy), 11 (10% moderate and 80% heavy), the MoE-UNI_CLAM model demonstrated improvement in AUC over UNI_CLAM_Base of 0.052, 0.019, and 0.047, respectively. Similar results were observed for MoE-CNN_simple and MoE-CNN_CLAM.

In contrast, in scenarios where the majority of tiles within a WSI were sharp or slightly blurred, e.g., scenarios 1, 9, and 12, all evaluated models, including both baseline and MoE approaches, achieved consistently high AUCs.

The UNI feature extractor consistently outperformed the ResNet18-based CNN feature extractor across all blur conditions, in both baseline model structure and MoE framework. Additionally, CLAM-based attention aggregation consistently improved performance relative to the 75th-percentile tile-to-slide aggregation for CNN_simple, underscoring the value of attention mechanisms for feature aggregation. This was particularly evident in challenging conditions such as scenario 11, where UNI_CLAM_Base significantly outperformed CNN_simple_Base (AUC = 0.861 vs. 0.533).

Table 4. Comparison of AUC scores between MoE and baseline models across 12 simulated WSI scenarios for NHG grading classification task (NHG 1 vs 3).

| Senarios | % of tiles within a WSI from Gaussian blur level L/M/H * | AUC | | | | | |
|---|---|---|---|---|---|---|---|
| | | MoE-CNN_simple | CNN_simple_Base | MoE-CNN_CLAM | CNN_CLAM_Base | MoE-UNI_CLAM | UNI_CLAM_Base |
| 1 | L = 100 | 0.886 | 0.886 | 0.901 | 0.901 | 0.952 | 0.949 |
| 2 | M = 100 | 0.808 | 0.733 | 0.868 | 0.702 | 0.950 | 0.928 |
| 3 | H = 100 | 0.746 | 0.421 | 0.828 | 0.493 | 0.879 | 0.578 |
| 4 | L/M/H = 50/25/25 | 0.870 | 0.867 | 0.900 | 0.893 | 0.948 | 0.939 |
| 5 | L/M/H = 25/50/25 | 0.843 | 0.810 | 0.890 | 0.875 | 0.944 | 0.931 |
| 6 | L/M/H = 25/25/50 | 0.823 | 0.800 | 0.887 | 0.871 | 0.934 | 0.920 |
| 7 | L/M/H = 50/50/0 | 0.865 | 0.867 | 0.902 | 0.892 | 0.951 | 0.943 |
| 8 | L/M/H = 0/50/50 | 0.788 | 0.658 | 0.862 | 0.611 | 0.929 | 0.877 |
| 9 | L/M/H = 80/10/10 | 0.883 | 0.887 | 0.903 | 0.901 | 0.952 | 0.947 |
| 10 | L/M/H = 10/80/10 | 0.823 | 0.749 | 0.878 | 0.834 | 0.950 | 0.931 |
| 11 | L/M/H = 10/10/80 | 0.782 | 0.533 | 0.858 | 0.790 | 0.908 | 0.861 |
| 12 | L/M/H = 80/15/5 | 0.883 | 0.886 | 0.903 | 0.900 | 0.952 | 0.947 |

*L, M and H represented varying proportions of low blur (L, $\sigma \in [0, 1.5]$), moderate blur (M, $\sigma \in (1.5, 5.0)$), and high blur(H, $\sigma \in [5.0, 10.0]$) tiles within a WSI across 12 simulation scenarios.

## Performance results in classification of IHC biomarker status from H&E status for top-performing models

To further evaluate prediction performance of the proposed MoE strategy, we performed benchmarking tasks of IHC biomarker status (i.e. ER, PR, Her2) predictions from H&E stained WSIs. Here we only consider the UNI_CLAM framework, based on the UNI feature extractor and attention-based aggregation, due to its superior performance over CNN alternatives (see previous section). Using the same training strategy, for each biomarker (ER, PR, and Her2), we evaluated the classification performance of one baseline model (trained on the original sharp images) and ten blur-resistant models each trained on a set of tiles added with a specific level of blur ($\sigma \epsilon$ {0.5, 1.0, 2.0, 3.0, 4.0, 5.0, 6.0, 7.0, 8.0, and 9.0}). We derived LV-based MoE model selection cut-offs following the same procedure described for grade classification (see Supplementary Figures S1–S3, and Table S6). We evaluated the model performance of the MoE approach (i.e., MoE-UNI_ER, MoE-UNI_PR, and MoE-UNI_Her2) compared to the corresponding baseline UNI_CLAM models (trained on unblurred tiles) under 12 simulated scenarios (Table 5). Across all three IHC prediction tasks, the MoE approach consistently performed better than the baseline models. In scenarios where WSIs consisted entirely of a single blur level (i.e. low, moderate, or high) (scenarios 1–3), the MoE approach outperformed the baseline models, with the largest improvement noted in scenario 3, where all tiles were heavily blurred. The same trend was also observed in mixed-blur scenarios 4–12, which more likely resemble the real-world WSI blur conditions. For instance, in scenario 9 (80% low, 10% moderate, and 10% high blurry tiles), which may reflect a common real-world scenario where a small proportion of tiles are of poor quality, MoE-UNI_CLAMs outperformed baseline models by margins of 0.017 (ER), 0.019 (PR), and 0.007 (Her2). Comparable improvements were also seen under the similar blur scenario 12 (80% low, 15% moderate, and 5% high blurry tiles).

Table 5. Comparison of AUC scores between MoE and baseline models across 12 simulated WSI scenarios for ER, Her2, and PR classification tasks.

| | | AUC | | | | | |
|---|---|---|---|---|---|---|---|
| **Senarios** | **% of tiles within a WSI from Gaussian blur level L/M/H *** | **MoE-UNI _CLAM_ER** | **UNI_CLAM_ ER_Base** | **MoE-UNI _CLAM_HER2** | **UNI_CLAM_ HER2_Base** | **MoE-UNI _CLAM_PR** | **UNI_CLAM_ PR_Base** |
| 1 | L = 100 | 0.917 | 0.903 | 0.773 | 0.766 | 0.763 | 0.758 |
| 2 | M = 100 | 0.912 | 0.874 | 0.766 | 0.734 | 0.762 | 0.713 |
| 3 | H = 100 | 0.845 | 0.559 | 0.691 | 0.550 | 0.691 | 0.524 |
| 4 | L/M/H = 50/25/25 | 0.916 | 0.892 | 0.770 | 0.756 | 0.768 | 0.718 |
| 5 | L/M/H = 25/50/25 | 0.916 | 0.881 | 0.762 | 0.743 | 0.763 | 0.695 |
| 6 | L/M/H = 25/25/50 | 0.900 | 0.864 | 0.756 | 0.727 | 0.751 | 0.657 |
| 7 | L/M/H = 50/50/0 | 0.916 | 0.899 | 0.778 | 0.756 | 0.771 | 0.748 |
| 8 | L/M/H = 0/50/50 | 0.897 | 0.834 | 0.744 | 0.694 | 0.745 | 0.620 |
| 9 | L/M/H = 80/10/10 | 0.916 | 0.899 | 0.771 | 0.764 | 0.765 | 0.746 |
| 10 | L/M/H = 10/80/10 | 0.914 | 0.879 | 0.763 | 0.738 | 0.764 | 0.706 |
| 11 | L/M/H = 10/10/80 | 0.868 | 0.792 | 0.723 | 0.659 | 0.718 | 0.582 |
| 12 | L/M/H = 80/15/5 | 0.918 | 0.901 | 0.770 | 0.762 | 0.766 | 0.753 |

*L, M and H represented varying proportions of low blur (L, σ∈[0, 1.5]), moderate blur (M, σ∈(1.5, 5.0)), and high blur(H, σ∈[5.0, 10.0]) tiles within a WSI across 12 simulation scenarios.

# Empirical assessment of compute time for the proposed MoE approach

Regarding time complexity, model training time for the MoE strategy scales linearly with the number of pre-defined blur levels, i.e., O(n), where n denotes the number of expert models (one per blur level). This time complexity is primarily influenced by the available computational resources, such as the number of GPUs, and is incurred only once during model development.

At inference, the time complexity per tile was O(1) for both sharpness evaluation and feature extraction by expert models, as each tile required only a single sharpness computation and straightforward routing to the appropriate expert. This constant-time complexity applies to all model architectures in this study.

To assess computational efficiency, we randomly selected 50 WSIs (including 63,118 tiles) from the validation set to be inferred under a simulated 100% moderate blur condition (scenario 2). Based on this sample data, we compared average inference time at WSI level between single base model approach and MoE approach. For each single baseline model, we noted time required for feature extraction and tile-to-slide aggregation, using either the 75th percentile or attention mechanism. For each MoE model, (i.e. MoE-CNN_simple, MoE-CNN_CLAM, and MoE-UNI_CLAM), we recorded time required for LV calculation, feature extraction from the selected experts, and tile-to-slide aggregation using either 75th percentile or the attention mechanism. We noted that feature extraction with UNI was slower than ResNet, reflecting its greater model complexity (Table 6). MoE models required more computation than baseline models due to the additional sharpness estimation and dynamic expert assignment, but the increase in mean inference time per slide remained modest and feasible for large-scale deployment.

Table 6*. Comparison of time complexity between baseline models with MoE models

| Baseline vs. MoE models | WSI average time(s) | Total time for the test set of 50 WSIs used for time estimation (s) |
|---|---|---|
| **CNN_simple vs. MoE-CNN_simple** | 2.640 vs. 5.049 | 131.883 vs. 252.472 |
| **CNN_CLAM vs. MoE-CNN_CLAM** | 2.106 vs. 4.576 | 105.303 vs. 228.772 |
| **UNI_CLAM vs. MoE-UNI_CLAM** | 13.706 vs. 16.186 | 685.290 vs. 809.288 |

* For MoE-CNN_CLAM and MoE-UNI_CLAM, the WSI average time included both LV estimation (2.478s) and model selection plus model inference time (0.016s for MoE-CNN_CLAM and 0.014s for MoE-UNI_CLAM). LV estimation during feature extraction for MoE models took a mean time of 1.961ms ± 0.082ms per tile. Then, the total time for LV estimation for all WSIs was 123.905s. The total time for assigning expert models and inference for all WSIs was 0.776s for MoE-CNN_CLAM and 0.683s for MoE-UNI_CLAM.

# Discussion

In this study, we assessed the impact of image blur on the performance of deep learning models for common breast cancer histopathology image analysis classification tasks, including histological grade prediction and prediction of IHC biomarkers from H&E histopathology images. To address the decreased performance caused by local unsharpness in WSIs, we proposed a MoE strategy with a sharpness-based gating mechanism. We found that unsharp images had a measurable impact on deep learning performance across all architectures evaluated, underscoring the importance of data collection and quality control of digital pathology images, as well as the importance of designing for robust AI solutions for histopathology analyses. Our primary contribution is that we have demonstrated that the proposed MoE approach enables models to actively handle WSIs with locally varying degrees of blur with improved performance, without eliminating large regions or entire slides, which has been a commonly used strategy previously.

In WSIs, it is not uncommon for some areas of the image to be less sharp than the rest. Using NHG classification as a benchmarking task, we showed that all evaluated model architectures (i.e. CNN_simple, CNN_CLAM, and UNI_CLAM) could be effectively trained to specialise in evaluating images with particular levels of blur. Based on this finding, we further implemented a MoE approach that selectively chooses the most suitable expert model for inference, depending on the blurriness of the input. We compared MoE with the generic single-model approach across 12 validation scenarios, each simulating a plausible blur pattern in WSIs in real-world situations. Our results showed that under these simulated mixed-blur scenarios, the MoE consistently achieved superior prediction performance compared to the single-model baselines. Of note, MoE-UNI_CLAM achieved strong performance (AUC ranging from 0.879 to 0.952) under a wide range of blur conditions, including the most challenging scenarios 3 and 11, where 80%-100% of tiles within a WSI were heavily blurred. Beyond NHG grade prediction, MoE also consistently outperformed single-model baselines on three additional tasks involving the prediction of ER, PR, and Her2 IHC biomarker status. Although MoE led to significant improvement in model performance, it only introduced a modest computational overhead from the additional step of blur estimation used for expert model selection. We have demonstrated that the proposed MoE approach can effectively enhance model prediction performance across varying image qualities, while maintaining time efficiency and scalability for large-scale applications.

Our analyses revealed that the addition of Gaussian blur led to a monotonic decline in the performance of baseline models trained with sharp images, with the decrease being more severe at higher blurring levels. One possible explanation is that the Gaussian blur tends to smooth images and reduce high-frequency features relevant to the prediction task at hand. On the other hand, blur-resistant models, trained on datasets with varying levels of blur, did not generalise uniformly across different blur conditions but often achieved the peak performance and outperformed the baseline models when evaluating data that matched the quality of the training set. These findings suggest that model parameters learned by blur-resistant expert models were better adapted to low-quality images, highlighting the limitation of deploying a single model for analyses across varying levels of image sharpness. This observation held true across all tested model architectures. When comparing baseline models of CNN_simple (using 75th-percentile aggregation), and baseline models of CNN_CLAM and UNI_CLAM incorporating attention-based tile-to-slide aggregation, we observed that models with attention modules generally outperformed CNN_simple across different blur scenarios. Interestingly, in scenario 11, one of the most challenging scenarios, with low/moderate/high blur tiles of 10%/10%/80%, CNN_CLAM and UNI_CLAM retained majority of the prediction power (AUC=0.790 and 0.861, respectively), significantly exceeding the performance of CNN_simple (AUC=0.533). This suggests that the attention-mechanism might be able to identify and prioritise sharp or low-blurred tiles despite the majority of tiles being severely degraded, or simply prioritise morphologies that are less sensitive to varying degrees of blur.

Previous studies have characterised how image quality affects CNN models[17,49], and proposed assessment tools for detecting the OOF regions in WSIs to mitigate its impacts[12,14,18,20]. For example, Kohlberger et al. developed ConvFocus [13] and Senaras et al. developed DeepFocus[50] to exhaustively localise and quantify the severity of OOF regions on WSIs. Rodrigues et al. generated an AI-based tool SlideQC[51] and Janowczyk et al. developed an open-source tool HistoQC[52] for automated quality control of WSIs. The blurred tiles or a highly blurred entire WSI detected by such approaches were then discarded. However, these methods primarily focused on quality control, typically followed by exclusion of low quality data, with exclusion criteria that may be inherently subjective. Furthermore, data exclusion may result in the loss of valuable information and may undermine the overall representativeness of the dataset. Alternatively, rescanning or re-sectioning and re-prepreparing tissue slides to gain higher quality images might not be practical at scale in routine digital pathology slides, particularly as many slides are partially impacted by local areas that are unsharp, and in practice, perfect image quality is hard to achieve. Therefore, a computationally efficient method capable of utilising lower-quality data offers a more effective alternative. We proposed a simple and efficient MoE framework, requiring only a one-time training of multiple expert models. Based on straightforward blur estimation of the input data using LV, it integrates predictions from multiple experts, effectively mitigating the impact of blurred images on deep learning model performance. The algorithm requires minimal computational overhead and can therefore be scaled up for large studies as well as being used in clinical diagnostic applications. We believe the proposed methodology can preserve valuable pathological information, which in turn could improve diagnostic accuracy in large-scale digital pathology studies and in real-world clinical use.

This study has some limitations. First, although the simulation study design was suitable for practical reasons, we were only able to cover a limited number of plausible scenarios. Second, we evaluated three commonly used deep learning model structures, and focused on four benchmarking tasks. Follow-up studies are warranted to investigate the impact of blur on a broader range of model architectures, and ideally, assess its impact across different applications. Another limitation, which could also be considered a strength, is that in the MoE implementation using the ViT foundation model as a feature extractor, only the attention-based model was optimised under different blur levels, not the feature extractor. We can imagine that potentially even higher performance could be gained if also the foundation model were re-optimised with image data of different blur levels, however, this would come with a computational overhead during training, and since results are highly promising, this might not be needed in practical applications. The strengths of this study include the population-based cohort design with a substantial number of WSIs, and the use of a simulation-based approach that offers a well-controlled context for both studying the impact of blur and evaluating the advantage of proposed novel MoE approach.

In conclusion, unsharp tiles reduce the performance of deep learning models in the context of digital and computational pathology. In clinical applications, including AI models for outcome prediction (e.g. prognostic or treatment response predictive applications), the possibility of biased or erroneous results for individual patients due to variable image quality could potentially lead to both over- and under-treatment. In such scenarios, improved management of scenarios with partially unsharp images might have a significant role to play to reduce such risks. This study highlights the importance of acquiring high-quality images, implementing robust quality control measures, and most importantly, exploring novel modelling strategies to mitigate issues associated with low-quality data. The proposed MoE approach improved performance under a range of conditions, and the approach may contribute to mitigating the impact of low image quality in the research context as well as in clinical applications.

# List of abbreviations

AI: artificial intelligence

H&E: hematoxylin and eosin

WSIs: whole slide images

CNNs: convolutional neural networks

ViT: Vision Transformer

CLAM: clustering-constrained-attention multiple-instance learning

FMs: foundation models

MoE: Mixture of Experts

NHG: Nottingham Histological Grade

IHC: immunohistochemistry

ER: estrogen receptor

PR: progesterone receptor

Her2: human epidermal growth factor receptor 2

LV: variance of Laplacian

CV: cross-validation

OOF: out-of-focus

AUC: area under the ROC curve

SGD: Stochastic Gradient Descent

LR: learning rate

# Acknowledgements

Not applicable.

# Funding

This work is supported by funding from the Swedish Cancer Society (20 0906 PjF 01 H; 23 2905 Pj 01 H), Karolinska Institutet (KID-funding), VINNOVA and SweLIFE (SwAIPP and SwAIPP2 projects),

MedTechLabs (BCAIND), Swedish e-science Research Centre (SeRC) (eCPC/eMPHasis project), and Swedish Research Council (2020-02163, 2022-01110). BL is supported by the Swedish Research Council (2019-06360).

# Data Availability

Data in the study cannot be deposited in a public repository and cannot be distributed without access control due to local privacy laws. Reasonable access requests to the corresponding author will be considered. All data analyses are based upon publicly available software packages (see Methods).

# Contributions

All authors contributed significantly to the work reported in this manuscript. YX contributed to method development, software implementation, validation, statistical analysis, writing the original draft of the manuscript, and editing the manuscript. BL contributed to the supervision, reviewing, editing, and approval of the manuscript. MR conceived the study, contributed to conceptualisation, resources, supervision, methodology, reviewing, editing, and approval of the manuscript, and directed the study.

# Competing interests

MR is a shareholder of Stratipath AB. All other authors have declared no conflicts of interest.